\documentclass[conference]{IEEEtran}
\IEEEoverridecommandlockouts
\usepackage{cite}
\usepackage{amsmath,amssymb,amsfonts}
\usepackage{graphicx}
\usepackage{algorithm}
\usepackage{algorithmicx}
\usepackage{algpseudocode}
\usepackage{multirow}
\usepackage{array}
\usepackage{subfigure}
\usepackage{enumerate}
\usepackage{textcomp}
\usepackage{url}
\usepackage{rotating}
\usepackage{adjustbox}
\usepackage{color}
\begin{document}

\title{External Service Sensing (ESS): Research Framework, Challenges and Opportunities}

\author{\IEEEauthorblockN{Zhongjie Wang, Mingyi Liu,  Zhiying Tu, Xiaofei Xu}
\IEEEauthorblockA{\textit{Faculty of Computing, Harbin Institute of Technology}\\
 Harbin, China \\
 \{rainy, liumy, tzy\_hit, xiaofei\}@hit.edu.cn}
 }

\maketitle

\begin{abstract}
The flourish of web-based services gave birth to the research area \textit{services computing},
a rapidly-expanding academic community since nearly 20 years ago. Consensus has been reached on a set of representative research problems in services computing, such as service selection, service composition, service recommendation, and service quality prediction. An obvious fact is that most services keep constant changes to timely adapt to changes of external business/technical environment and changes of internal development strategies. However, traditional services computing research does not consider such changes sufficiently. Many works regard services as \textit{static} entities; this leads to the situation that some proposed models/algorithms do not work in real world. Sensing various types of service changes is of great significance to the practicability and rationality of services computing research. In this paper, a new research problem \textit{External Service Sensing} (ESS) is defined to cope with various changes in services, and a research framework of ESS is presented to elaborate the scope and boundary of ESS. This framework is composed of four orthogonal dimensions: sensing objects, sensing contents, sensing channels, and sensing techniques. Each concrete ESS problem is defined by combining different values in these dimensions, and existing research work related to service changes can be well adapted to this framework. Real-world case studies demonstrate the soundness of ESS and its framework. Finally, some challenges and opportunities in ESS research are listed for researchers in the services computing community. To the best of our knowledge, this is the first time to systematically define service change-related research as a standard services computing problem, and thus broadening the research scope of services computing.
\end{abstract}

\begin{IEEEkeywords}
Services Computing, Research Framework, External Service Sensing, Challenges and Opportunities
\end{IEEEkeywords}

\section{Introduction}\label{sec:intro}
With the rapid development of SOA, IoT, Cloud and Edge computing, the number of available services has exploded and the types of services has become increasingly diverse. 
The flourish of web-based services gave rise to the research area called \textit{services computing}, and its academic community has kept continuously expanding since nearly 20 years ago\cite{wu2014service}.
Services computing involves many classical research problems, such as service selection\cite{kayastha2015survey}, service discovery\cite{mukhopadhyay2012survey}, service composition\cite{sheng2014web}, service classification\cite{pang2019augmenting}, Quality of Service (QoS) prediction\cite{ghafouri2020survey}, and service recommendation\cite{zhu2018similarity}.


An obvious fact is that most real-world services, both web-based services and business services, keep constant changes to timely adapt to changes of external business/technical environment and changes of internal competitive strategies. Service providers actively or passively make adjustments to their services. The types of changes include technical-level QoS and functionalities and business-level contents and performances such as price, promotion policies, and delivery channels; new services may enter the market, and dying services may exit. However, traditional services computing research does not consider service changes sufficiently. Many research works regard services as \textit{static} entities; consequently, resulting in the situation that many proposed models/algorithms do not work in the real world. For example, for the service classification problem, when a service adds or remove certain features, the category it belongs to may change and the original classification results are no longer accurate; and for service recommendation, ignoring service changes may result in recommending a service that no longer conforms to the user's demand or preferences, or even worse, recommending a service that has been deprecated.
It is necessary to summarize such kinds of issues related to service changes and define a consensual research problem to deal with these formidable challenges that service changes present to traditional services computing problems. In this paper, we called it \textit{External Service Sensing} problem (ESS). The keyword \textit{sensing} indicates that research on this problem is to perceive service changes timely and accurately. The keyword \textit{external} is emphasized because this problem stands on the view of external observers of service to be sensed. Users of the service usually do not have the privilege to access the internal status of the service due to legal, business, ethical, and technical barriers, but they have to perceive the changes from the outside of the service. A significant challenge is that external observers perceive service changes in terms of the limited information that is exposed by the providers of the service, and to make the sensing more timely and accurately, new and effective approaches have to be elaborately studied.

Note that to sense a service from the internal point of view has been widely studied in the past, such as run-time service performance monitoring\cite{taisch2014service,he2020optimal}. This falls out of the scope of ESS and this paper.

\begin{figure}
    \centering
    \includegraphics[width=\linewidth]{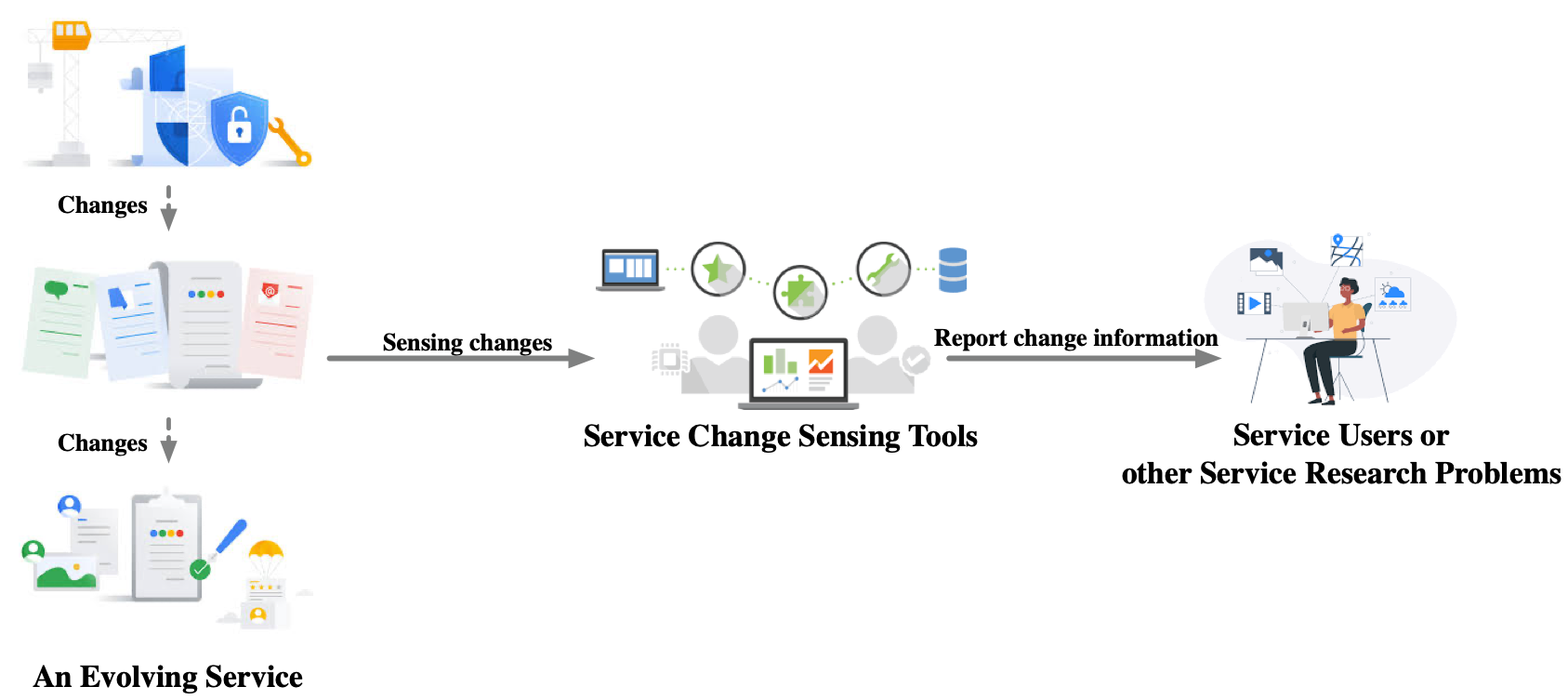}
    \caption{Illustration of the External Service Sensing (ESS) problem}
    \label{fig:intro}
\end{figure}


Fig.~\ref{fig:intro} briefly demonstrates the ESS problem. More specifically, an ESS problem could be decomposed into the following dimensions:
\begin{enumerate}
    \item What are the objects to be sensed? This is called ``sensing objects''.
    \item What kinds of perspectives of a sensing object are expected to change and need to be sensed? This is called ``sensing contents''.
    \item What necessary data are required for the sensing and where are these data collected? This is called ``sensing channels''.
    \item How are concrete change information of sensing contents identified from sensing channels? This is called ``sensing techniques''.
\end{enumerate}

There are many options in each of the four dimensions, and a concrete ESS problem is therefore defined by the combination of different strategies adopted in these perspectives. 

The Benefits of a systematic study on the ESS problem are twofold. Firstly, incorporating timely and accurate sensing results into traditional services computing problems would drive services computing researchers to invent new models and methods for these problems so that the practicability and rationality could be significantly improved. Secondly, the presence of ESS would help identify many new application scenarios in services computing research. For example, service providers could adjust their business strategies in time by sensing changes in competitors' services. Regulators could develop appropriate policies to guide the healthy development of a service market by comprehensively sensing market changes. Most of these scenarios have not yet received sufficient attention from the services computing community.

\begin{figure*}
    \centering
    \includegraphics[width=\textwidth]{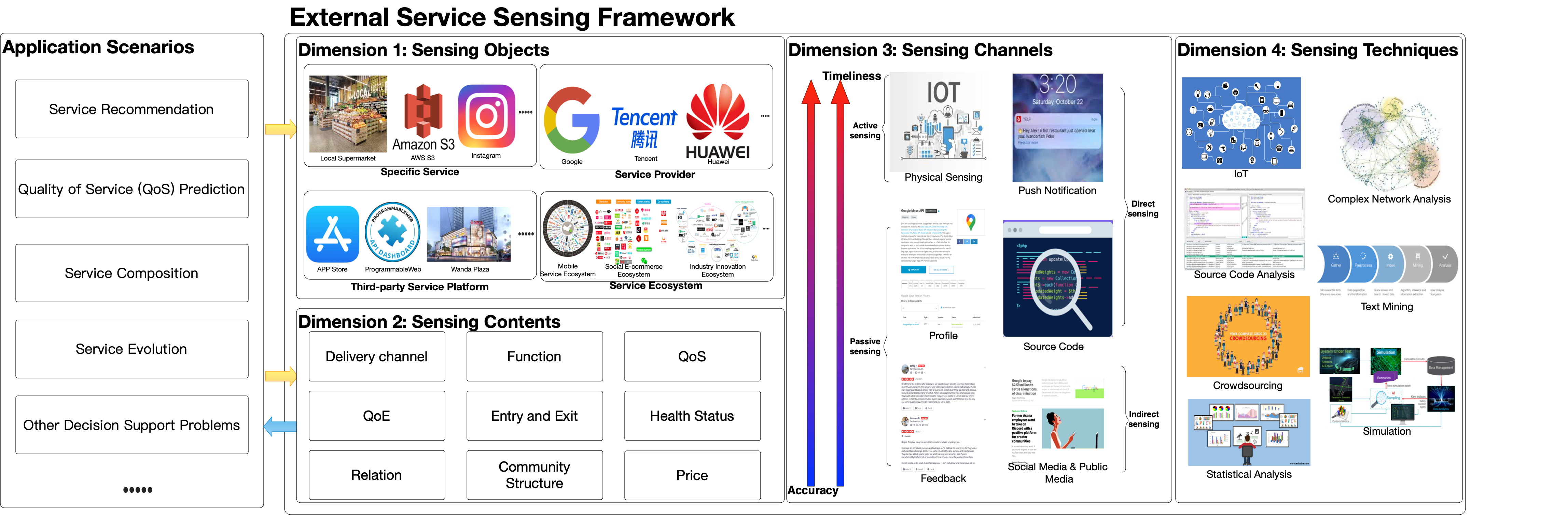}
    \caption{A four-dimensional framework for the External Service Sensing (ESS) problem}
    \label{fig:framework}
\end{figure*}

In recent years, there have been some studies involving specific perspectives of service changes. For example, Zhong et al.\cite{zhong2014time} presented a method for perceiving changes in the invocation relations between APIs and mashups in a web service ecosystem for more accurate service recommendation; Huang et al.\cite{huang2017time} gave an approach for sensing QoS fluctuations in IoT environments to rank available services; Liu et al.\cite{liu2020community} proposed a method for predicting the service ecosystems' evolutionary trends by tracking the changes of service communities; and Hao et al.\cite{huang2017time} analyzed the changes of invocation relations between Apps by parsing source codes of these Apps. The most immediate work is by Zheng et al.\cite{zheng2010distributed,zheng2012investigating}, which introduced the distributed infrastructure to access Web services periodically and evaluate their QoS so that the fluctuations of QoS could be perceived timely.

Each of these studies can be considered as a concrete ESS problem, indicating the exploration of ESS has already been launched sporadically in the services computing community. However, since ESS has not yet been widely recognized as an independent services computing problem, these existing works usually treat the service sensing task as a subsidiary of other services computing problems. This greatly lowers the significance and challenges of ESS. To compensate for this, we are committed to a systematic discussion of ESS and propose a four-dimensional research framework to elaborate the scope and boundary of ESS. The main contributions of this paper are listed as follows:

\begin{itemize}
    \item  To the best of our knowledge, this is the first time that the problem of ESS is explicitly and systematically studied. We explore the definition of ESS, the relevance of ESS to other traditional services computing problems, and the significance that ESS would bring to the services computing community.
    
    \item A comprehensive research framework of ESS is proposed, which defines the four dimensions (sensing objects, sensing contents, sensing channels, and sensing techniques) of ESS and the relations between them. It lays a solid foundation for other researchers to carry out ESS research in the future. Concrete ESS problems could be easily instantiated from this framework.
    
    \item We collect some representative existing ESS related works from services computing literature and use the framework to classify them. This demonstrates that the framework is compatible with existing research. We also summarize some challenges and opportunities for future research if ESS is widely accepted as a standard service research problem by the community.
\end{itemize}

The remainder is organized as follows. Section \ref{sec:framework} introduces the proposed framework in detail. Section \ref{sec:case_study} illustrates the effectiveness of the framework by systematic mapping and case studies. Section \ref{sec:vision} gives some challenges and opportunities of ESS. The last section is the conclusion.

\section{An ESS Research Framework}\label{sec:framework}
As mentioned in Section~\ref{sec:intro}, ESS is an upstream task of other services computing problems, such as service recommendation, QoS prediction, and service composition. Here we call them \textit{application scenarios} of ESS. The input of an ESS problem are the sensing objects and sensing contents required by a specific application scenario, the output is the information of changes of sensing contents, and the solution of the ESS problem is to use specific sensing techniques to collect data via specific sensing channels and make data analysis to get the output which is further utilized in the application scenario.

Fig.~\ref{fig:framework} shows the ESS framework which contains four internal dimensions, namely \textbf{sensing objects}, \textbf{sensing contents}, \textbf{sensing channels} and \textbf{sensing techniques}. The first two dimensions focus on \textbf{what} to be sensed, the third on \textbf{where} to sense, and the fourth on \textbf{how} to sense. Subsequent part of this section delineate each dimension respectively, and the relations between them.

\subsection{Dimension 1: Sensing Objects}\label{sec:objects}
In terms of sensing granularity, sensing objects can be classified into specific services, service providers, third-party service platforms, and service ecosystems:

\begin{itemize}
    \item \textbf{Specific service}: A service is delivered and provisioned by an organization or individuals to offer specific functionalities for meeting specific user demands. It may be a software service (e.g., a mobile App \textit{Instagram}, a web API \textit{Amazon S3 API}) or a physical facility in the real world (e.g., an \textit{Amazon Go} store). Since external observers have no chance to access the interior of service, services are the finest sensing unit in ESS. 

    \item \textbf{Service provider}: including service companies (e.g., \textit{Google}, \textit{Alibaba}) and government regulators ( e.g., \textit{State Post Bureau of China}). Sensing a service provider is usually accompanied by sensing the specific services it offers. Thus it has broader application scenarios. For example, users would like to know the latest released services of a company, and a company could timely adjust its business strategies according to the sensing of its competitors.
   \item \textbf{Third-party service  platform (TPP)}: A TPP is operated by a specific organization and aggregates massive service providers and the corresponding services they offer. Users may select and use appropriate services from a large number of candidates. A TPP can be an online platform such as \textit{App Store}, \textit{ProgrammableWeb}, an offline physical platform such as \textit{Wanda Plaza}, or a platform linking online and offline such as an e-commerce platform \textit{Amazon.com} and a takeaway platform \textit{Meituan}. Sensing TPP includes sensing the entry and exit of services or service providers to/from the TPP, ranking of services on the TPP, and so on. 
   
   \item \textbf{Service ecosystem}: It is a type of complex continuously-evolving systems consisting of service providers and specific services that belong to specific regions or domains. Compared with TPP, service ecosystems typically have a larger size and cover a wider range of services and service providers. There are no explicit operators responsible for the provision of a service ecosystem (it evolves spontaneously). For example, the e-commerce service ecosystem in China is composed of thousands of online e-commerce websites, mobile Apps, offline marts, and billions of potential customers. Sensing changes in service ecosystems is often used in scenarios that support decision-making for service providers as well as market regulators.
\end{itemize}

It should be noted that there is no strict boundary between these four categories of sensing objects, and multiple sensing objects may be simultaneously involved in an ESS application scenario.

\subsection{Dimension 2: Sensing Contents}
Sensing contents are elements or perspectives of a specific sensing object, and they might change along with time due to external or internal drivers. 

There are many types of sensing contents. For example, functionalities of a service may be expanded or compressed, Quality of Service (QoS) and Quality of Experience (QoE) may be improved or deteriorated, price of service may increase or cut down, and relations between two services may emerge or disappear. Different types of sensing objects may have a different sets of sensing contents.

Representation forms of sensing contents are diverse. Some sensing content could be represented by continuous and precise values, such as QoS and price. Some are discrete and text-based, e.g., functions and delivery channels. There are also some sensing contents that are subjectively measured, such as QoE and health status. 

The degree of attention paid to different sensing contents varies among existing studies: some sensing contents have been well-studied, such as QoS. Community structures, relations, and functions are emerging sensing contents in recent studies. Note that existing studies have focused more on the static metrics of sensing contents than on their changes.

\subsection{Dimension 3: Sensing Channels}\label{sec:channels}
Sensing channels are the location where necessary data related to the changes of sensing objects and sensing contents are acquired. Sensing channels can be classified into the following six types:

\begin{itemize}
    \item \textbf{Physical sensing} refers to the acquisition of sensing objects and contents from offline physical space and then transfer data to online cyberspace. Physical sensing may rely on hardware sensors, e.g., flight tracking\footnote{https://www.flightaware.com/live/}, smart home\cite{jiang2018smart}, and smart city\cite{mirri2014combining,9399954}. Physical sensing may also rely on volunteers (persons) to report changes they manually perceive, but obviously this approach requires significant labor costs, e.g., local traffic condition reporting\cite{anciaes2019perceptions} or crime incident reporting\cite{mohler2018privacy}. This is also called ``social sensing''.
    
    \item \textbf{Push notification} means that a sensing object actively publicizes its changes to whom it may concern. This channel is widely used in reality, such as promotional emails, App update notifications, etc.
    
    \item \textbf{Profiles} are text descriptions of a sensing object and are usually publicized by the sensing object itself, e.g., the description text of API service. Profiles have been widely used in a variety of services computing researches on account of the progress of natural language process techniques, but few studies have focused on sensing changes from service profiles.
    
    \item \textbf{Source code}. On the one hand, with the prevalence of open source communities, many services expose their source codes to the public, e.g., \textit{Telegram}\footnote{https://github.com/DrKLO/Telegram}. On the other hand, the distribution format of some types of software services has been standardized. However, these services do not provide publicly-accessed source codes, some configuration files used to describe services are publicly accessible. For example, an Android application file \textit{APK} always contains a file called \textit{AndroidManifest.xml}. Changes of such types of services may be partially sensed via the updates of their source codes or configuration files.

    \item \textbf{User feedback} refers to the evaluation of a sensing object from its users. Feedback can be a quantitative score such as the rating of Apps in the \textit{App Store}, or text-based descriptions such as user reviews on \textit{Yelp}. Changing trends of users' sentiments on specific perspectives of service is a valuable type of change to be sensed.

    \item \textbf{Social media \& public media} refer to acquire changes of sensing objects from the discussion contents in public social networking sites or public news reports. For example, there is a great deal of information about \textit{Windows} on \textit{Reddit}, and a wealth of investment events among service providers can be found from investment information sites\footnote{https://www.crunchbase.com}. Similar to the user feedback channel, change-related information in social media \& public media is represented by the changing trend of opinions from massive users or the public. Compared with other sensing channels, this channel offers richer information about changes, and the cost for obtaining change information is comparatively lower. However, irrelevant and fake news would affect the accuracy of the sensing results.
    
    \begin{figure*}[thbp]
    \centering
    \includegraphics[width=\textwidth]{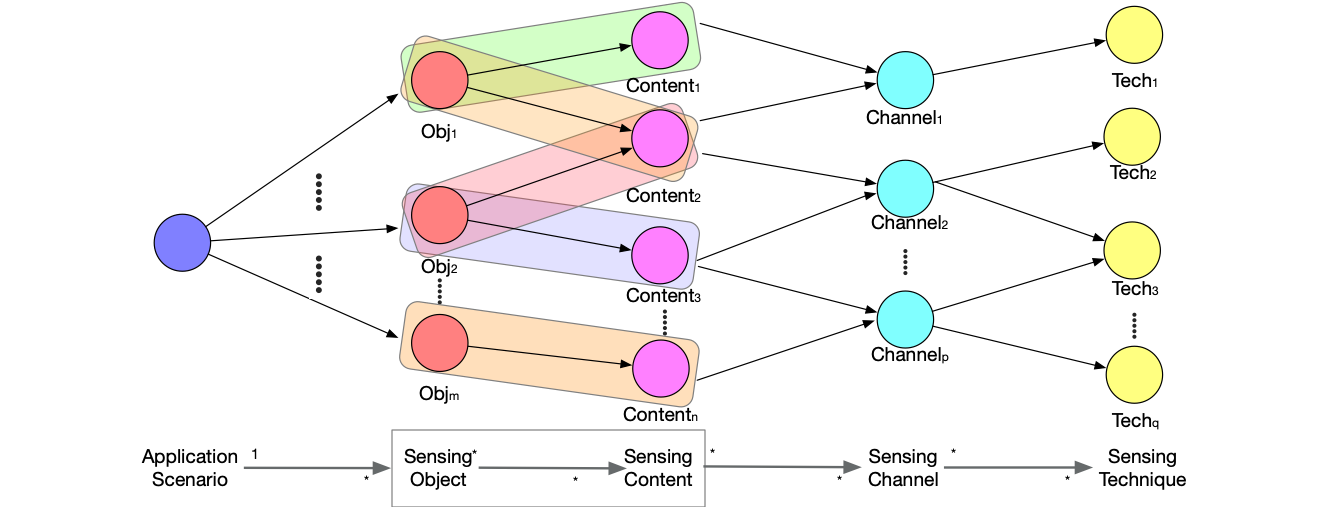}
    \caption{Relations between the four dimensions}
    \label{fig:relation_of_dim}
\end{figure*}
\end{itemize}

The above-mentioned sensing channels could be classified into active and passive ones in terms of whether the change information is actively and directly publicized by the sensing objects or not. Physical sensing and push notification belong to active sensing channels, and others are passive ones. Compared with passive sensing channels, active sensing could more timely provide more accurate information about changes, but the sensing contents are often limited and incomplete. This drives researchers to pay more attention to passive ones to pursue more comprehensive sensing.

From another perspective, sensing channels can also be classified into direct and indirect ones depending on whether the sensing data is directly acquired from a sensing object or other sources related to the sensing object. Feedback and social media \& public media are indirect sensing channels, while others are direct ones. The sensing latency of direct sensing channels would be lower than that of indirect ones, i.e., the timeliness of sensing is higher. Indirect channels exist more widely, but as mentioned above, there are lots of noisy data that require elaborate processing.

\subsection{Dimension 4: Sensing Techniques}
Sensing techniques are processing methods for extracting changes of sensing contents from the data acquired from sensing channels.

Some sensing techniques are tied closely to specific sensing channels. For example, IoT techniques utilize hardware sensors for physical sensing, and source code analysis techniques parse the source code of service to obtain specific changes of the sensing contents. 

Some sensing techniques have been widely used in other services computing problems and can be easily transferred to ESS, e.g., crowdsourcing approaches allow volunteers to actively report changes of sensing objects and contents, and data-driven deep learning methods such as text mining are used to extract changes from a user feedback or public media texts. These methods facilitate sensing changes from a massive amount of cheap data efficiently and accurately.

Sensing techniques are not independent of each other. Some techniques are based on the data generated by other sensing techniques before they can process to obtain the target sensing contents. For example, complex network analysis relies on the graph constructed by other sensing techniques to obtain sensing contents such as community structure and health status of a service ecosystem, while statistical analysis methods aggregate explicit sensing contents obtained by other sensing techniques to calculate implicit sensing content that is not directly available.

\begin{table*}[]
\caption{Adaptation of some existing works to the proposed ESS framework}\label{tab:summary}

\begin{adjustbox}{width=\textwidth} 
\begin{tabular}{lccccc}
\hline
                                                         & \begin{tabular}[c]{@{}c@{}}Application\\ Scenario\end{tabular} & \begin{tabular}[c]{@{}c@{}}Sensing \\ Objects\end{tabular}                               & \begin{tabular}[c]{@{}c@{}}Sensing\\ Contents\end{tabular}                 & \begin{tabular}[c]{@{}c@{}}Sensing\\ Channels\end{tabular}           & \begin{tabular}[c]{@{}c@{}}Sensing\\ Techniques\end{tabular}                                                   \\ \hline
Tu et al. \cite{tu2021bidirectional}                     & App recommendation                                             & Mobile app                                                                               & Function                                                                   & Profile                                                              & Text mining                                                                                                    \\ \hline
Shu et al. \cite{shu2020social}                          & Service recommendation                                         & Real-world service                                                                       & \begin{tabular}[c]{@{}c@{}}QoS\\ Price\end{tabular}                        & \begin{tabular}[c]{@{}c@{}}Feedback\\ Public media\end{tabular} & Text mining                                                                                                    \\ \hline
Zheng et al. \cite{zheng2020price}                       & Service recommendation                                         & Items                                                                                    & \begin{tabular}[c]{@{}c@{}}Price\\ Category\end{tabular}                   & Profile                                                              & \begin{tabular}[c]{@{}c@{}}Source code analysis\\ Statistical analysis\end{tabular}                            \\ \hline
Zhong et al.\cite{zhong2014time}                         & Service recommendation                                         & \begin{tabular}[c]{@{}c@{}}Third-party service platform\\ (ProgrammableWeb)\end{tabular} & Invocation relation                                                        & Profile                                                              & Complex network analysis                                                                                       \\ \hline
Hao et al. \cite{hao2016global}, \cite{hao2016empirical} & Service evolution                                              & \begin{tabular}[c]{@{}c@{}}Mobile app,\\ Mobile service ecosystem\end{tabular}           & \begin{tabular}[c]{@{}c@{}}Interface\\ Function\\ Cooperation\end{tabular} & Source code                                                          & \begin{tabular}[c]{@{}c@{}}Source code analysis\\ Complex network analysis\\ Statistical analysis\end{tabular} \\ \hline
Liu et al. \cite{liu2020community}                       & Service evolution                                              & Business service ecosystem                                                               & Community structure                                                        & Public media                                                    & \begin{tabular}[c]{@{}c@{}}Text mining\\ Complex network analysis\end{tabular}                                 \\ \hline
Song et al. \cite{song2020method}                        & Service evolution prediction                                   & Mobile app                                                                               & \begin{tabular}[c]{@{}c@{}}Function\\ QoE\end{tabular}                     & \begin{tabular}[c]{@{}c@{}}Profile\\ Feedback\end{tabular}           & Text mining                                                                                                    \\ \hline
Adeleye et al. \cite{adeleye2018constructing}            & Service discovery                                              & \begin{tabular}[c]{@{}c@{}}Third-party service platform\\ (ProgrammableWeb)\end{tabular} & Relation                                                                   & Profile                                                              & Complex network analysis                                                                                       \\ \hline
Xia et al. \cite{xia2019efficient}                       & Service discovery                                              & IoT services                                                                             & \begin{tabular}[c]{@{}c@{}}Relation\\ (Correlation degree)\end{tabular}    & Physical sensing                                                  & \begin{tabular}[c]{@{}c@{}}Complex network analysis\\ Simulation\end{tabular}                                  \\ \hline
Huang et al.\cite{huang2017time}                         & Service ranking                                                & IoT services                                                                             & QoS                                                                        & \multicolumn{1}{l}{Physical sensing}                              & \begin{tabular}[c]{@{}c@{}}Simulation\\ Statistical analysis\end{tabular}                                      \\ \hline
\end{tabular}\end{adjustbox}
\end{table*}
 
\subsection{Relations between the Four Dimensions}
The four dimensions in the framework are not independent but are correlated with each other. Relations between them are shown in Fig.~\ref{fig:relation_of_dim}.

For an application scenario, several different sensing objects may need to be sensed, and these sensing objects can be of different granularity. A sensing object can contain more than one element/perspective, and the ones relevant to the application scenario are selected as the sensing contents. Different sensing objects may own some common sensing contents. Since different sensing contents of a sensing object are often distributed in different sensing channels, it is reasonable to bundle sensing objects and sensing content to support the application and find appropriate sensing channels. The selection of appropriate sensing techniques is often based on the form of data provided by the sensing channel.

Generally speaking, the dimensions of the ESS framework have relatively obvious upstream and downstream relations, and different dimensions are with many-to-many relations.

\section{Systematic Mappings and Case Studies}\label{sec:case_study}
In this section, we first demonstrate a mapping from some existing work related to service changes to the proposed framework, i.e., to re-define the problems in these related work using ESS concepts. The objective is to show the rationality of the framework, i.e., it is compatible with previous services computing research. Then we give some case studies to show that this framework has the capability of defining real-world scenarios into standard ESS problems, i.e., it has good practicality.

\subsection{Systematic Mappings}

Table~\ref{tab:summary} shows the adaptation of some existing works to the proposed ESS framework. Although these works originally were not  intended to address the ESS problems, we can see that each of them can be represented by the instantiation of the ESS framework. For example, the goal of Tu et al.\cite{tu2021bidirectional} is to make mobile service recommendations, the object of their sensing is mobile Apps, and text mining are the sensing technique to obtain functions changes of Apps from App profiles. With the help of the sensing results of function changes of Apps, more reasonable and accurate recommendation results could be obtained than other App recommendation approaches without considering sensing service function changes. Due to limited space, we do not explain other works any more.

Note that the ESS framework would inspire services computing researchers to extend their work by incorporating change-related information into their work. The framework gives researchers a systematic thinking way for this change-oriented extension. For example, Hao et al.\cite{hao2017service} perform service recommendation by analyzing service descriptions without considering dynamic entry and exit of services, which leads to many deprecated services in the recommendation results. The recommendation of deprecated services could be avoided if such dynamic changes were sensed. This problem can be solved by simulating user requests to check the availability of services during the recommendation periodically.

\subsection{Case Studies}

\begin{figure*}[thbp]
    \centering
    \includegraphics[width=\textwidth]{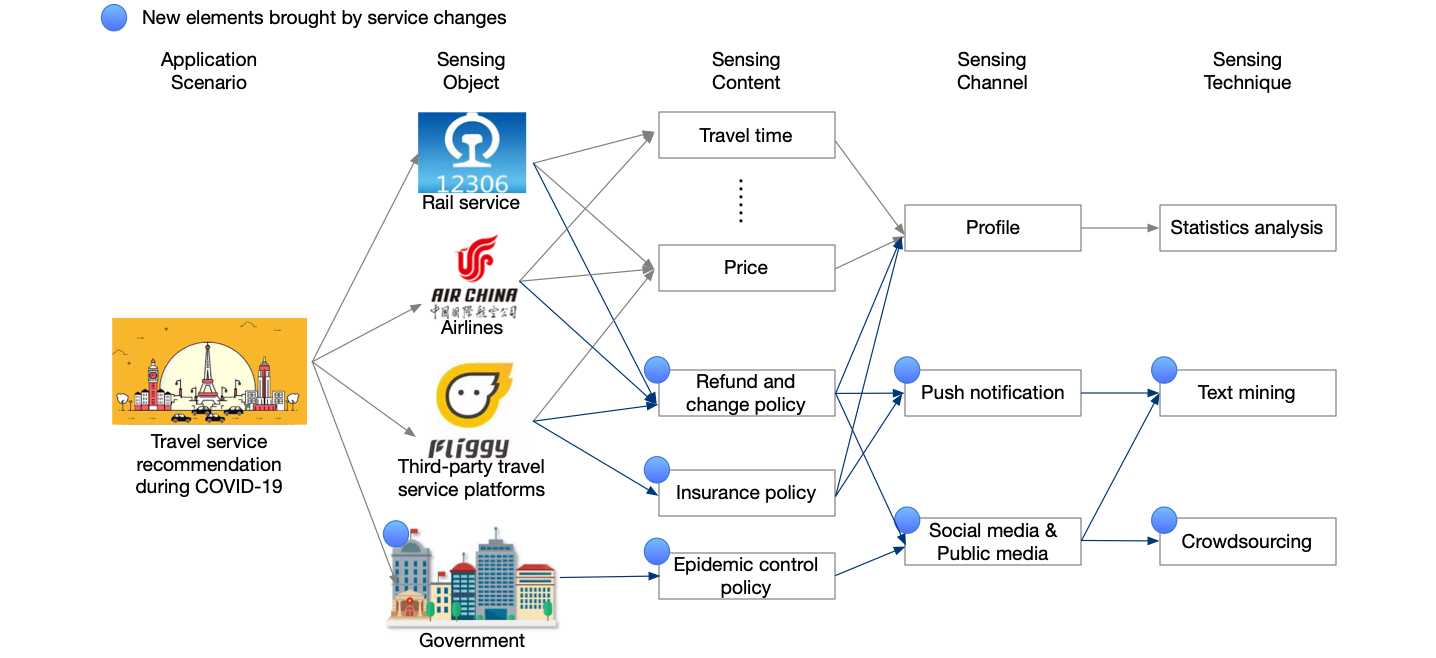}
    \caption{A case study: travel service recommendation during COVID-19 pandemic}
    \label{fig:case}
\end{figure*}

Here we give a real application scenario about travel service recommendation during COVID-19. It is then defined as a concrete ESS problem by the framework.

During COVID-19, customer travel planning will be affected by various uncertainties, such as epidemic control policy around the world, and refund, change, and insurance policies of airlines and rail services. There are very frequent changes occurring in these services, and it is impracticable to recommend a travel plan to customers without considering these changes. For example, some cities may unexpectedly deny access to people from regions with high COVID-19 infection risk. Therefore, it is necessary to sense those changes that might occur on travel-related services and incorporate them into the recommendation.

Fig.~\ref{fig:case} gives the demonstration of using our ESS framework to define such travel service recommendation problem. The sensing objects include: (1) travel service providers and third-party travel platforms and the travel services they offer; (2) as governments play an important role in epidemic prevention and control, they also need to be sensed. For sensing contents, besides the common QoS attributes of the travel services, e.g., price, travel time, additional COVID-19 related frequently-changing perspectives such as refund and change policies, insurance policies, and epidemic policies, should be sensed to reduce the risk of travel, too. These sensing contents can be obtained from sensing channels such as service profiles, governments/services' push notifications, and published media. Since most of the information is disclosed to the public in the form of text, text mining is the sensing technique that has to be adopted. And statistical analysis and crowdsourcing are essential supplements. The sensing results are then incorporated into the travel service recommendation method to select the most appropriate travel route and flights that suit the preferences of a specific user, e.g., minimizing the risk of refund and cost.

Another case that has been widely used in the real world is driving route planning. The route with the shortest distance is usually recommended. But when traffic condition changes are considered, the shortest route may not be the fastest one. When considering the external service sensing, text mining may be applied to acquire continuously updated road maintenance information from the local government's push notification (e.g., websites and mobile Apps); crowdsourcing and IoT sensing are also employed to sense the traffic flow directly from the physical world (this has been widely adopted in many mobile navigation Apps such as Baidu Map). Dynamic traffic condition information is then transferred to route planning algorithms to recommend a route for avoiding congestion and shortening actual driving time.
    
ESS could also support the decision-making of service market regulators. Taking the bike-sharing service ecosystem as an example, the number of service providers and bikes they delivered to the market, the quality of bikes, the price per ride, and so on, keep dynamically changing over time. Market regulatory authority needs to monitor the development health status of this ecosystem and timely make appropriate policies to guide service delivery strategies of bike-sharing companies or the investment behaviors in stock markets. From the ESS point of view, the sensing object is a specific service ecosystem, the content is the health status of the ecosystem, and complex network analysis and statistical analysis could be used to identify the changes of various industry development metrics from transaction/investment data of various related companies. This case has been elaborately studied in 
Liu et al previous work\cite{liu2020data}.

From the three cases we can see that ESS problems exist widely in real world, and this demonstrates the potential of ESS research and the value of ESS applications. It is worthy to put more effort into ESS research and applications.

\section{Challenges \& Opportunities}\label{sec:vision}
Although a framework has been proposed to define ESS problems systematically, there are still many challenges to meet. Here we list some critical challenges for services computing researchers as research opportunities.

\subsection{Challenge 1: How to standardize the description and release channels of service changes?}

As we mentioned in Section~\ref{sec:framework}, there are different granularities of sensing objects, ranging from small specific services to large service ecosystems. The forms of sensing contents are diverse, too, ranging from numerical values to textual information. It is important to standardize the representation of these heterogeneous changes so that others could access and use the change-related information in a simple and convenient way. One possible solution is to create a Service Change Agreement (SCA) standard, just like using Service Level Agreement (SLA)\cite{jin2002analysis} to standardize the commitment between a service provider and a client covering particular aspects of the service such as quality, availability, and responsibilities. How to publicize SCA of massive services to reduce the cost of duplicate sensing is also a challenging problem. A third-party SCA aggregation platform that provides standardized APIs for publicizing and acquiring SCAs might be a solution.

\subsection{Challenge 2: How to measure the quality of sensing?}

Inaccurate, incomplete, and delayed sensing would mislead the downstream tasks (recommendation, prediction, etc.) to unacceptable output, and too costly sensing is hard to accept neither. Designing metrics for ESS tasks is an urgent challenge and of great significance to form a standard evaluation system for future ESS solutions. In our opinions, four metrics are essential:

\begin{enumerate}
    \item Accuracy: to what extent the sensing results could reflect the changes occurring in services? In other words, the perceived change-related information should be inconsistent with what has happened as far as possible. It is similar to the ``precision'' in machine learning.
    \item Comprehensiveness: i.e., the changes that have occurred in services should be perceived as many as possible. It is similar as the ``recall'' in machine learning. 
    \item Timeliness: how long is it between the time when changes occurred and the time when these changes have been perceived? The longer the time lag is, the lower the timeliness of sensing is, and vice versa. Outdated change information is meaningless.
    \item Cost: refers to how much it needs to pay to conduct the sensing and get results, including the cost for change-related data collection and processing. Of course, users would like to perceive changes with lower cost.
\end{enumerate}

We do not present the detailed measurement of these metrics. Different application scenarios have different levels of tolerance on them. 

\subsection{Challenge 3: How to protect privacy during sensing?}

Similar to user privacy, sensing objects also have their privacy. For example, a service provider hopes that its latest discount policies are known only to a limited number of its loyal users but should not be sensed by others (such as its competitors). 

To solve this problem, one possible idea is to mimic the \textit{robots} protocol used in web crawlers to indicate what types of changes are allowed or not to sense. Even for those changes that are allowed to sense, service objects may have the willingness to control the dissemination scope of these changes, and new solutions have to be studied to manage the privacy issues comprehensively.

Additionally, frequent sensing may increase the burden on the sensing objects, e.g., simulating user requests would increase the network burden of the service, and sensing objects can reduce this burden by proactively publishing its changes through the third-party SCA aggregation platform proposed in Challenge 1.

\subsection{Challenge 4: How to sense offline services more efficiently and cost-effectively?}

As we mentioned in Section~\ref{sec:channels}, physical sensing channel relies on hardware sensors or crowdsourcing (\textit{social sensors}), which are much more costly than online service sensing. To alleviate this problem, cheaper sensors and more widespread deployment of sensors are required, and more efficient crowdsourcing strategies are to be further explored. On the other hand, novel ways of virtualizing offline services into cyberspace need further study. Synchronizing offline worlds with online worlds is considered as a big challenge. Fortunately, more and more practices have been conducted, such as the industrial Internet that facilitates the creation of \textit{digital twins} of the two worlds.

\subsection{Challenge 5: To build ESS infrastructure and toolset}

It remains a challenge to implement our proposed framework from a technical level. A set of tools and infrastructure that support the whole sensing process is required, such as the SCA aggregation platform and sensing-related APIs mentioned in Challenge 1. Such infrastructure and tools should provide common processing flows of ESS and enhance the sensing capabilities of service application; on the other side, they are to reduce the burden of developers who previously have to develop the sensing functionalities, so that developers are empowered by out-of-the-box built-in tools for different ESS dimension. The architecture and operational logic of this platform may be similar to that of current machine learning platforms, such as Amazon SageMaker\footnote{https://aws.amazon.com/cn/sagemaker/}.

\section{Conclusion}\label{sec:conclusion}
In this paper we define a new research problem \textit{External Service Sensing} based on our observations of service change related research and practice. We present a framework for unifying related work and guiding future research. We make a systematic mapping from existing service sensing related work to the proposed framework and validates the effectiveness of the framework. Some cases from real world also demonstrate that ESS problems indeed widely exist and need more attention. We also discuss some challenges of ESS that have not yet been fully defined and solved, which would be a \textit{uncultivated land} of services computing research. We expect this paper would attract more researchers from services computing community to denote themselves to ESS problems.



\bibliographystyle{IEEEtran}
\bibliography{reference}

\end{document}